\documentclass[reprint,  pra, superscriptaddress]{revtex4-1}
\usepackage{amsmath}
\usepackage{amssymb}
\usepackage{amsfonts}
\usepackage{braket}
\usepackage{graphicx}

\newcommand{\be}{\begin{equation}}
\newcommand{\ee}{\end{equation}}

\newcommand{\mc}[1]{\mathcal{#1}}

\newcommand{\fig}[1]{Fig.~\ref{#1}}
\newcommand{\figs}[2]{Figs.~\ref{#1} and \ref{#2}}
\newcommand{\Fig}[1]{Figure~\ref{#1}}

\newcommand{\eq}[1]{Eq.~\eqref{#1}}

\newcommand{\stn}[1]{Sec.~\ref{#1}}

\newcommand{\bea}{\begin{eqnarray}}
\newcommand{\eea}{\end{eqnarray}}
 
\newcommand{\ba}{\begin{array}}
\newcommand{\ea}{\end{array}}

\newcommand{\bl}{\begin{flalign}}
\newcommand{\enl}{\end{flalign}}

\begin{document}
\title{Stark Control of Electrons Across the Molecule-Semiconductor Interface} 

\author{Antonio J. Garz\'on-Ram\'irez}
\affiliation{Department of Chemistry, University of Rochester, Rochester, New York 14627, United States}
\altaffiliation{Present Address: Department of Chemistry, Northwestern University, Evanston, Illinois 60208-3113, United States}
\author{Ignacio Franco}
\email{ignacio.franco@rochester.edu}
\affiliation{Department of Chemistry, University of Rochester, Rochester, New York 14627, USA}
\affiliation{Department of Physics, University of Rochester, Rochester, New York 14627, USA}
\date{\today}

\begin{abstract}
Controlling matter at the level of electrons using ultrafast laser sources represents an important challenge for science and technology. Recently we introduced a general laser control scheme (the Stark Control of Electrons at Interfaces or SCELI) based on the Stark effect that uses the subcycle structure of light to manipulate electron dynamics at semiconductor interfaces [{Phys. Rev. B \textbf{98}, 121305 (2018)}]. Here, we demonstrate that SCELI is also of general applicability in molecule-semiconductor interfaces. We do so by following the quantum dynamics induced by  non-resonant few-cycle laser pulses of intermediate intensity (non-perturbative but non-ionizing) across model molecule-semiconductor interfaces of varying level alignments. We show that SCELI induces interfacial charge transfer regardless of the energy level alignment of the interface and even in situations where charge exchange is forbidden via resonant photoexcitation. We further show that the SCELI rate of charge transfer is faster than those offered by resonant photoexcitation routes as it is controlled by the subcycle structure of light.  The results underscore the general applicability of SCELI to manipulate electron dynamics at interfaces on ultrafast timescales.
\end{abstract}

\maketitle

\section{Introduction}

 Characterizing and controlling matter driven far from equilibrium represents a major challenge for science and technology.\cite{brumerbook, ricebook, Mukamel_book, Heller_2018, Tannor_2007, Lienau_2020, Scholes_2017, Cavalleri_2016, Cavalleri_2011, ivanov2009,  Corkum2007, Cavalleri_2007, Kohler_2005,    Marangos_2005, Butikov_2001, bergmann, hanggiCDT,   Paul_1990, Mollow, AutlerTownes}
A particularly important goal is to design methods for the control of matter at the level of electrons with ultrafast laser sources. The reason to focus on electrons is because they determine the chemical reactivity, optical and transport properties of matter. The reason to focus on lasers as a source of control is because they allow for the manipulation of matter on ultrafast --femto to attosecond-- time scales, something that is not achievable by conventional means such as  applied voltages, thermodynamic or chemical control.

The latest advances in laser technology now enable the generation and control of few-cycle lasers in the IR and UV/Vis with well-defined carrier envelope phase. Using such pulses it is now possible to apply laser fields with intensities of $\sim 10^{13}-10^{14}$ W cm$^{-2}$ (amplitudes of $\sim 1-2$ V/\AA) before the emergence of dielectric breakdown. At those intensities, the incident light can dramatically distort the electronic structure of nanoscale systems and bulk matter as the strength of the light-matter interaction becomes comparable to the strength of chemical bonds, thus opening unprecedented opportunities to manipulate  electronic properties and dynamics on ultrafast time scales.  Strong field effects that are expected at such electric field amplitudes include Zener interband tunneling of electrons,\cite{Zener1934} Bloch oscillations,\cite{Bloch1929, Hartmann_2004} the appearance of localized electronic states and the associated Wannier-Stark ladder in the energy spectrum,\cite{Wannier1960, Schockley, DiCarlo1994, Leitenstorfer2018} the emergence of Floquet replicas and low-energy absorption features in the optical absorption spectrum\cite{Gu2018_2,Kobayashi2023}, 
and the response of electronic dynamics to the subcycle structure of light.\cite{franco2007,liping_2018,Antonio_2020,Antonio_2018,Boolakee2022,doi:10.1021/acs.nanolett.1c02538,Higuchi2016,Schultze2013,Schiffrin2013,Schiffrin2016}

\begin{figure}[htbp]
\centering
\includegraphics[width=0.5\textwidth]{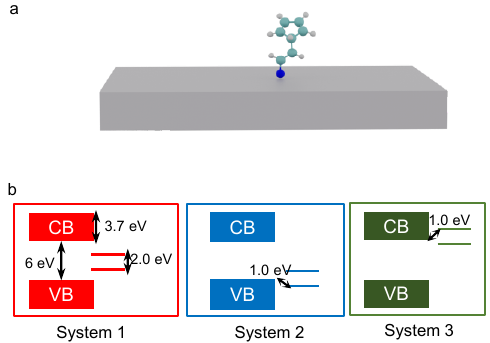}
\caption{Schematic of the molecule-semiconductor interface. (a) Sketch of the interface and (b) the different energy level alignments \label{fig:fig1}}
\end{figure}

In particular, few-cycle laser pulses enable the generation of very large Stark effects\cite{krausz1998, Sussman5ways, AutlerTownes, Fukuyama,franco2007,liping_2018,Antonio_2018,Antonio_2020,Schiffrin2013,Zener1934,Schultze2013,Leitenstorfer2018,Hartmann_2004}  
that effectively modify the energy levels of the system in a dynamic fashion during the duration of the pulse, creating transient laser-dressed materials with effective properties that can be very different from those observed near thermodynamic equilibrium. For example, through Stark effects it is now possible to bridge the 9 eV energy gap of fused silica (glass) and transiently turn this dielectric into a material akin to a  metal.\cite{Schultze2013,Schiffrin2013}  Because at these field strengths the Stark effect depends on the instantaneous value of the electric field (as opposed to the pulse envelope), few cycle laser also offer control of the dynamics of level alignment simply by changing the shape of the laser --by modifying its carrier envelope phase.

In this context, we recently proposed a general laser control scheme to manipulate electron dynamics at interfaces, the Stark Control of Electrons at Interfaces or SCELI~\cite{Antonio_2018,Antonio_2020,Antonio_2020_2},  that is based on Stark effects triggered by few cycle lasers. SCELI uses non-resonant lasers of intermediate intensity (non-ionizing but whose effect cannot be captured by finite-order perturbation theory)   to distort the electronic structure of matter through Stark shifts and create 
 transient resonances between the  electronic levels of two adjacent materials $A$ and $B$. When the level of material $A$ is occupied and the one in $B$ empty, these transient crossings  open  quantum tunneling pathways for interfacial $A\to B$ charge transfer.
 We have demonstrated how to use SCELI to transiently turn  an insulating semiconductor-semiconductor heterojunction into a conducting one,\cite{Antonio_2018} to induce phase-controllable currents in spatially symmetric systems,\cite{Antonio_2020, liping_2018} and its robustness to screening and  band-bending effects.\cite{Antonio_2020_2}  Further, since SCELI is a form of Hamiltonian control based on shifting the energy levels of matter, we find that SCELI  is robust to decoherence \citep{Antonio_2018,Antonio_2020,Antonio_2020_2} since it does not rely on creating electronic superposition states with fragile coherence properties. This is important since electronic decoherence in matter is remarkably fast (typically in $\sim$ 10s fs) \cite{doi:10.1063/1.4984128,doi:10.1021/acs.jpclett.7b01817,doi:10.1021/acs.jpclett.7b03322,Wen_2018} limiting the applicability of interference-based laser control schemes\cite{Batista2002,Rego2009,doi:10.1063/1.2940796}.
 Thus far, SCELI has been computationally demonstrated in semiconducting interfaces \citep{Antonio_2018,Antonio_2020} and to be at play in laser-induced rectification in gold-silica-gold nanojunctions \citep{liping_2018}.
 
As a novel frontier, here we extend the domain of applicability of SCELI to the  molecule-semiconductor interface. 
Molecule-semiconductor or molecule-metal interfaces are of interest because they determine the operation of molecular electronics \citep{Mangold2011,Jafri2015,doi:10.1021/acs.nanolett.5b00877,doi:10.1021/acs.chemrev.5b00680,doi:10.1021/acs.jpclett.7b03323,Galperin_2007, ZIMBOVSKAYA20111,doi:10.1063/1.5003306}, STM imaging \cite{Cocker2016,Bartels648,PhysRevLett.119.176002,Czap670,Peller2020}, catalysis \citep{PhysRevLett.122.077401,Jiang2013,KOCH2018194}, electrochemistry \citep{Zhou1998,DENBOER2016390,Grill2020} and other surface phenomena. In particular, the ultrafast control of electron dynamics at the STM-molecule interface now enables studies of electron transfer at the level of single electron \citep{ PhysRevB.82.085444,Jelic2017} and the imaging of nuclear motions at the single molecule limit \cite{Cocker2016,Bartels648,PhysRevLett.119.176002,Peller2020}.
From the perspective of laser control, what makes this interface distinct is that it is atomically sharp  and that the Stark response of the molecule is significantly weaker with respect to that of extended systems \citep{Schindler2006,doi:10.1021/jp8067393,PhysRevB.81.125209}.

The remainder of this paper is organized as follows. In \stn{stn:model_and_methods} we define the model and laser pulses employed for control, and the method used to solve the quantum dynamics in this problem. In \stn{stn:results}, we discuss the SCELI mechanics, its magnitude and directionality in representative molecule-semiconductor interfaces. We also characterize the dependence of the effect on laser and molecular parameters, and contrast SCELI to traditional routes to induce charge transfer in molecule-semiconductor interfaces by resonantly exciting the semiconductor. We summarize our main observation in \stn{stn:conclusions}.

\section{Model and Methods}
\label{stn:model_and_methods}
\subsection{Hamiltonian}

A schematic representation of a molecule-semiconductor interface is shown \fig{fig:fig1}a. Here this interface is {modeled as a one-dimensional chain that grows along the  direction $\hat{i}$ normal to the interfacial plane, and is} represented through a tight-binding model with Hamiltonian
\begin{equation}
\label{eq:hamilt_total}
\hat{H}(t)=\hat{H}_{\text{S}}(t)+\hat{H}_{\text{M}}(t)+\hat{H}_{\text{MS}},
\end{equation} 
where $\hat{H}_{\text{S}}$ is the Hamiltonian for the semiconductor S, $\hat{H}_{\text{M}}$ that of the molecule M, and $\hat{H}_{\text{MS}}$ their interfacial coupling. In the model both the molecule and the semiconductor interact with a laser field $\mathbf{E}(t)=E(t) \hat{i}$ in dipole approximation {with polarization direction}  $\hat{i}$. 
The semiconductor is modeled through a two-band tight-binding Hamiltonian
\begin{equation}
\label{eq:tight_hamil}
\hat{H}_{\text{S}}(t)= \sum_{n=1}^{2N}(h^{\text{S}}_{nn}+|e|E(t)x_n)\hat{a}^{\dagger}_{n}\hat{a}_{n}+\sum_{\langle n,m\rangle}^{2N}h^{\text{S}}_{nm}(\hat{a}^{\dagger}_{n}\hat{a}_{m}+ \text{H.c.}),
\end{equation} 
where $\langle n,m\rangle$ denotes nearest-neighbors, and H.c. Hermitian conjugate.  Here $\hat{a}_{n}$ ($\hat{a}^{\dagger}_n$) annihilates (creates) a fermion in site or Wannier function $n$ and satisfies the usual fermionic anticommutation relations. Each unit cell has two Wannier functions with alternating on-site energies ($h^{\text{S}}_{nn}=h^{\text{S}}_{\text{even}}\delta_{n,\text{even}}+h^{\text{S}}_{\text{odd}}\delta_{n,\text{odd}}$) in tight-binding coupling among them ($h_{n,n+1}^{\text{S}}={t^{\text{S}}}$). In turn, $x_n$ denotes the position of each Wannier function along the junction and $|e|$ the electron charge. The molecule is described as a two-level system  with Hamiltonian 
\begin{equation}
\label{eq:molecu_hamil}
\hat{H}_{\text{M}}(t)=\sum_{n=2N+1}^{2N+2}({h^{\text{M}}_{nn}}+|e|E(t) x_n)\hat{a}^{\dagger}_{n}\hat{a}_{n}+V(\hat{a}^{\dagger}_{2N+1}\hat{a}_{2N+2}+ \text{H.c.})
\end{equation} 
where {$h^{\text{M}}_{nn}$} is the energy for molecular sites $n=2N+1$ and $n=2N+2$,  and $V$ the coupling between sites. Further, we assume that the semiconductor and the molecule interact through that molecular site that is attached to the surface such that $\hat{H}_{\text{MS}}=-t_{\text{MS}}(\hat{a}_{2N}^{\dagger}\hat{a}_{2N+1}+\text{H.c.}),$ where $t_{\text{MS}}$ is the interfacial tight-binding coupling.

For definitiveness, as a representative lattice constant for a semiconductor we choose $5.0$ \AA\ and distance between sites in each cell of 1.7~\AA{, such that $x_n = 5.0(n-1)/2~ \text{\AA}$ for $n\in \textrm{S}$ odd and  $x_n = (5.0 (n-2)/2 + 1.7) ~\text{\AA}$ for $n\in \textrm{S}$ even}. The tight-binding coupling ${t^{\text{S}}}=-3.0$ eV and the on-site energies $h^\text{S}_{\text{odd}}=0.0$\ eV,\ $h^\text{S}_{\text{even}}=6.0$ eV were chosen to have a semiconductor with a 6 eV bandgap and 3.7 eV bandwidth. We model the semiconductor with $N=50$ unit cells. The molecular parameters were chosen to yield the energy alignments show in Fig. \ref{fig:fig1}b with a molecular energy gap of 2 eV. The interfacial distance is $a_{\text{MS}}=$4.0 \AA\ and interfacial tight-binding coupling $t_{\text{MS}}=0.2$ eV.  

{We focus on molecules with and without permanent dipoles. For molecules with no net dipoles, $V=1.0$ eV and  ${h^{\text{M}}_{2N+1,2N+1}=h^{\text{M}}_{2N+2,2N+2}}=$ 3.0, 0.0 and 6.0 eV  for systems 1, 2 and 3, respectively.} 
The ground and excited molecular orbital eigenenergies are $\epsilon_\text{g/e}=h^{\text{M}}_{2N+1,2N+1} \pm |V|$. The oscillator strength 
$f_{\text{OS}}=\frac{4}{3}\frac{m_{e}}{|e|^2 \hbar^2}V |\mu_\text{eg}|^2$ (where $m_{e}$ the electron mass) is determined by the transition dipole $\mu_\text{eg}$ which can be manipulated by varying the distance between sites. 
{For molecules with permanent dipoles, we now have that  $h^{\text{M}}_{2N+1,2N+1}\ne h^{\text{M}}_{2N+2,2N+2}$. In this case,  $\epsilon_\text{g/e}=\overline{E} \pm\sqrt{\Delta E^2+4|V|^2}/2$, with $\overline{E}=(h^{\text{M}}_{2N+1,2N+1}+h^{\text{M}}_{2N+2,2N+2})/2$ and $\Delta E=h^{\text{M}}_{2N+1,2N+1}-h^{\text{M}}_{2N+2,2N+2}$. The transition dipole $\mu_\text{eg}=\sigma_\text{e}\sigma_\text{g}(x_{2N+1}+ x_{2N+2}\chi_\text{e}\chi_\text{g})$  {and the net molecular dipole $\mu_{jj}=x_{2N+1}\sigma_{j}^2+x_{2N+2}\chi_{j}^2\sigma_{j}^2$}, where $\sigma_j=1/\sqrt{1+\chi_j^2}$ and  $\chi_j=(h^{\text{M}}_{2N+1,2N+1}-\epsilon_j)/V$ ($j=$g,e). In practice, we choose $\epsilon_\text{g/e}$ and the oscillator strength $f_{\text{OS}}$ to match the case without net dipoles. This, together with a chosen value for the net dipoles $\mu_{ee} = -\mu_{gg}$ completely determines the parameters of this Hamiltonian. }

\subsection{Laser Pulse\label{sec:laser}}
The laser pulse employed in the simulations is a few-cycle laser of central frequency $\omega$, width $\tau=\sqrt{2}\pi/\omega$, centered around $t_c=50$ fs, and carrier envelope phase (CEP) $\phi$. The vector potential associated with the laser pulse used in the simulations is of the form 
\begin{equation}
\label{eq:vector_potential}
\mathbf{A}(t)=\frac{E_0}{\omega}{\rm e}^{-(t-t_c)^2/2\tau^2}\sin(\omega(t-t_c)+\phi)\hat{i}.
\end{equation} 
The associated electric field $\mathbf{E}(t)=-\frac{d\mathbf{A}(t)}{dt}=E(t)\hat{i}$ is given by 
\begin{multline}
\label{eq:laser}
E(t)=\frac{E_0}{\omega}{\rm e}^{-(t-t_c)^2/2\tau^2} \Big[\frac{(t-t_c)}{\tau^2}\sin(\omega(t-t_c)+\phi) \\ 
 -\omega\cos(\omega(t-t_c)+\phi)\Big].
\end{multline}
This form guarantees that $E(t)$ remains as an ac source even for few-cycle lasers as $\int_{-\infty}^{\infty}E(t)dt=0$. A few-cycle laser is chosen to suppress the onset of dielectric breakdown \cite{Schultze2013,krausz1998,Schubert2014} even for moderately strong fields with intensity $\sim 10^{13}-10^{14}$ W/cm$^2$. The laser frequency  is chosen to be far detuned from electronic transitions to suppress multiphoton absorption via near resonant photoexcitation. {Throughout, unless specified otherwise, we choose a laser frequency $\hbar\omega = 0.5$ eV and  CEP $\phi=0$. }

\subsection{Equation of Motion}

Since the Hamiltonian in Eq. \ref{eq:hamilt_total} is an effective single-particle operator, all the electronic properties are determined by the single-particle electronic reduced density matrix $\rho_{n,m}(t)=\bra{\Psi(t)}\hat{a}_n^{\dagger}\hat{a}_m\ket{\Psi(t)}$, where $\ket{\Psi(t)}$ is the many body wavefunction. The dynamics of $\rho_{n,m}(t)$ is governed by the Liouville-von Neumann equation \begin{equation}
    \label{eq:dynamics}
    i \hbar \frac{d}{dt}\rho_{n,m}(t)=\langle[\hat{a}_n^{\dagger}\hat{a}_m,\hat{H}]\rangle,
\end{equation} with initial condition $\rho_{nm}(0)=\sum_{\varepsilon=1}^{\mc{N}}\langle\varepsilon|n\rangle\langle m|\varepsilon\rangle p(\varepsilon)$, where $\ket{\varepsilon}$ are the single-particle eigenstate of the Hamiltonian of Eq. \ref{eq:hamilt_total} at time $t=0$, and $p(\varepsilon)$ the initial electronic distribution function. {We consider charge neutral systems for which the number of electrons $\mc{N}=N+1$} Equation \ref{eq:dynamics} is numerically integrated for a time of  $t=1000$ fs, using the predictor-corrector Adams-Moulton method with adaptive time step~\cite{sundials}. Snapshots of the key observables are recorded every $\Delta t_{\text{obs}}=0.01$ fs. 

 The electron transfer dynamics is monitored through changes in the total charge $\Delta Q_i$ of each material. Specifically, we focus on the net charging of the semiconductor $\Delta Q_{\text{S}} (t)=|e|\sum_{n=1}^{2N}(\rho_{nn}(t)-\rho_{nn}(0))$ and the molecule $\Delta Q_{\text{M}} (t)=|e|\sum_{n=2N+1}^{2N+2}(\rho_{nn}(t)-\rho_{nn}(0))$. Any $\text{S}\to\text{M}$ electron transfer is monitored by $Q_{\text{S}\to\text{M}}(t)=\Delta Q_{\text{M}} (t)$. The quantity $Q_{\text{S}\to\text{M}}(t)> 0$ when charge flows from S$\to$M, and negative when M$\to$S charge transfer is favored. The asymptotic charge $Q_{\text{S}\to\text{M}}(\infty)$ is recorded at a time $t=800$ fs well after the pulse decays.

\section{Results and Discussion}
\label{stn:results}

\subsection{SCELI mechanism}
\label{stn:SCELI}
\begin{figure}[htb]
   \centering
    \includegraphics[width=0.5\textwidth]{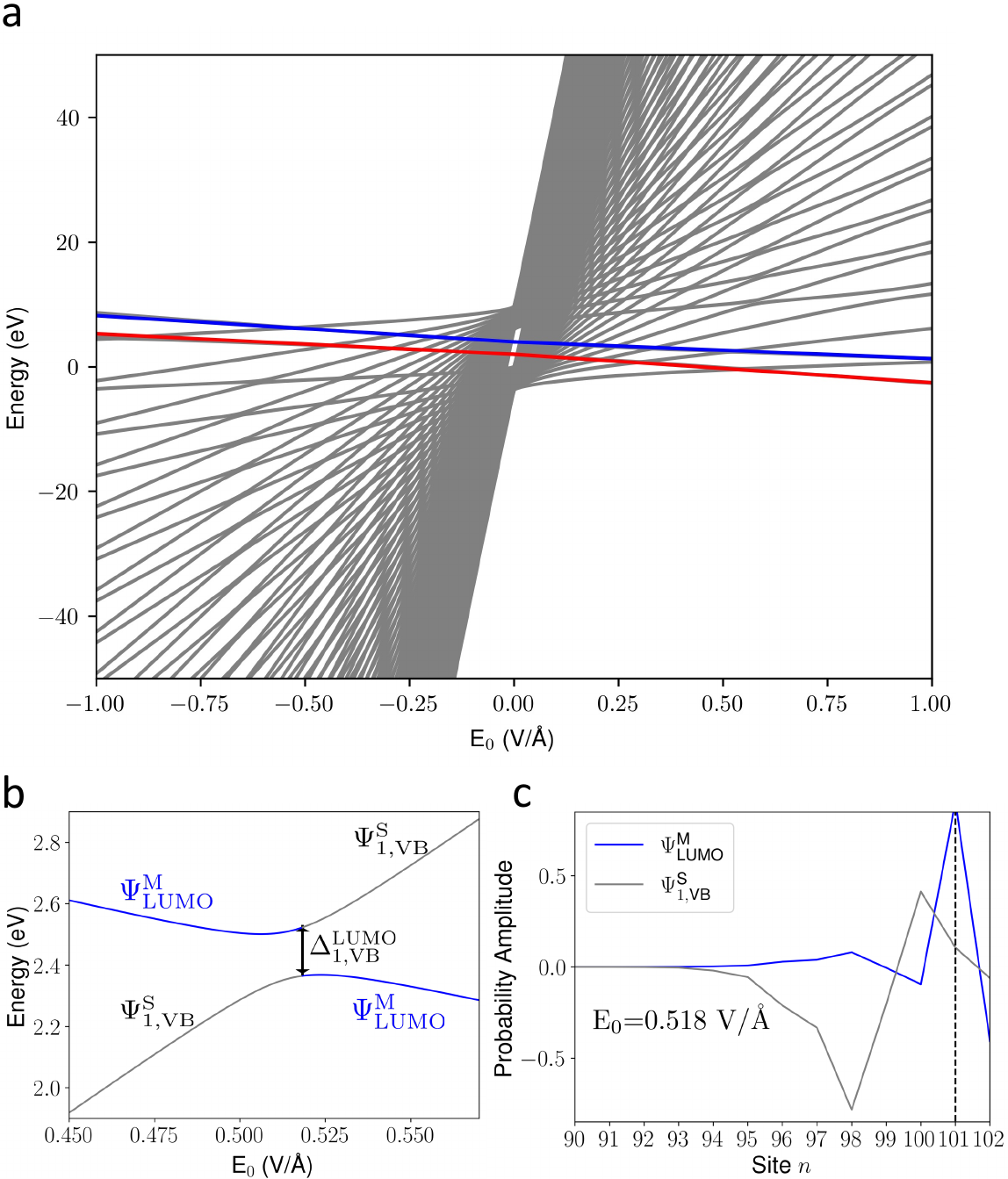}
    \caption{Electronic structure of the Semiconductor-Molecule interface in the presence of a static electric field $E_0$.  (a) Under the influence of an electric field the eigenenergies of the heterojunction (obtained by diagonalizing $\hat{H}$ [\eq{eq:hamilt_total}] for a fixed electric field $E_0$) fan out as the electric field amplitude $E_0$ increases resulting in multiple trivial and avoided crossings. The gray lines represent the semiconductor levels, the colored lines the molecular orbitals (red: HOMO, blue: LUMO). (b) Detail of an avoided crossing between molecular and semiconductor levels that open tunneling pathways for electron transfer. The crossing is between a VB level of S (with diabatic eigenfunction $\Psi_{1, \text{VB}}^{\text{S}}$) and the molecular LUMO ($\Psi^\text{M}_{\text{LUMO}}$). These crossing are particularly effective when the wave functions of the dressed  states spatially overlap at the interface ($n=$100-101, in this case) as in (c) leading to significant energy gap $\Delta$ at the avoided crossing. Parameters are for System 1 in \fig{fig:fig1}b. \label{fig:fig2}}
\end{figure}

To understand SCELI in the molecule-semiconductor interface it is useful to consider what happens to the electronic structure of a semiconductor and a two-level molecule in the presence of a static electric field $E_0$. 
\Fig{fig:fig2} shows the energy levels of the heterojunction for varying $E_0$, using System 1 in \fig{fig:fig1}b as an example.  When the semiconductor interacts with an electric field, its electronic structure is distorted through Stark shifts. This distortion destroys the periodicity of the potential of the semiconductor \citep{Stey_1973,GLUCK2002103,Hartmann_2004}. As a consequence the energy spectrum shows equally spaced resonances known as the Wannier-Stark ladder, and the wavefunctions become localized. This is best appreciated in a tight-binding one-band model \citep{Stey_1973, GLUCK2002103,Hartmann_2004} where the energy of the Wannier-Stark states are given by 
$\varepsilon_{m}=\varepsilon_{\text{S}0}+|e|maE_0$,
 where $\varepsilon_{\text{S}0}$ denotes the center of the energy band for the field-free model, $m=\cdots,-2,-1,0,1,2\cdots$ the site, $a$ the lattice constant. That is, the energy levels of each band of the semiconductor fan out due to the Stark shifts. Additionally, because the electric field introduces a linear potential, with a $|e|aE_0$ drop between consecutive sites, the wavefunctions become localized. The degree of localization increases as the electric field, and thus the potential drop between consecutive sites, increases.  Similarly, when the two-level molecule interacts with a static electric field its energy level repel due to Stark shifts with an energy level difference $2 |\mu_{\text{eg}} E_0|\sqrt{1+ \left(\frac{V}{\mu_{\text{eg}} E_0} \right)^2}$ where $2|V|$ is the energy level difference in the absence of the electric field and $\mu_{\text{eg}}$ is the transition dipole between the levels. Note that the Stark response of the molecule is small  with respect to that observed in the semiconductor as its reduced size limits the magnitude of $\mu_{\text{eg}}$, and that it becomes linear in $E_0$ for strong electric fields for which  $(V/\mu_{\text{eg}} E_0)^2 \ll 1$.

As a result of these distortions, resonances between the electronic diabatic eigenstates of the semiconductor $\text{S}$ and the molecule $\text{M}$  are formed at specific values of $E_0$. Most of these crossings will be trivial as there is no significant spatial overlap between the wavefunctions involved. However, when the wavefunctions of the two diabatic states overlap at the interface such as the crossing detailed in \fig{fig:fig2}c, this opens quantum tunneling channels for interfacial electron transport. These overlaps lead to an avoided crossing between the adiabatic energy levels of the system (eigenstates of the full Hamiltonian $\hat{H}$) such as those in \fig{fig:fig2}b.
When a valence band (VB) level of $\text{S}$ enters into transient resonance with the LUMO of $\text{M}$, $\text{S}\to \text{M}$ tunneling electron transfer across the heterojunction can occur. Similarly, when the crossing is between a conduction band (CB) level of $\text{S}$ and the HOMO, $\text{M}\to \text{S}$ tunneling electron transfer can occur.  

More explicitly, for a given fixed electric field $E_0$ the diabatic basis is defined by 
\begin{equation}
\label{eq:diabatic_wf}
 \hat{H}_\alpha\ket{\Psi^\alpha_{j}}=\varepsilon_{j}^{\alpha}\ket{\Psi_{j}^{\alpha}},
\end{equation} 
where $\hat{H}_\alpha$ is the Hamiltonian for material $\alpha=\text{S}, \text{M}$ in the presence of such electric field. In this basis, the Hamiltonian of the two levels involved in a given crossing is 
\begin{equation}
\hat{H}=
\begin{pmatrix}
\varepsilon_{i}^{\text{S}}&\Delta/2\\
\Delta/2&\varepsilon_{j}^{\text{M}}
\end{pmatrix},
\end{equation} where $\Delta/2=\bra{\Psi_i^{\text{S}}}\hat{H}_{\text{SM}}\ket{\Psi_{j}^{\text{M}}}=\bra{\Psi_j^{\text{M}}}\hat{H}_{\text{SM}}\ket{\Psi_i^{\text{S}}}$ is the coupling between the diabatic states. During a crossing, the diabatic states hybridize to yield adiabatic states (eigenstates of $\hat{H}$) 
\begin{equation}
\begin{aligned}
\ket{\Psi_+}&=-\sin\xi \ket{\Psi_j^{\text{M}}}+\cos\xi\ket{\Psi_i^{\text{S}}},\\
\ket{\Psi_-}&=\cos\xi \ket{\Psi_j^{\text{M}}}+\sin\xi\ket{\Psi_i^{\text{S}}}
\end{aligned}
\end{equation} 
with $\sin 2\xi=\Delta/\sqrt{\Delta^2+(\varepsilon_{i}^{\text{S}}-\varepsilon_{j}^{\text{M}})^2}$. This hybridization leads to the opening of an energy gap $\Delta$ among the adiabats and tunneling of population between the two diabatic states of the two materials during the crossing.

%The splitting between the adiabatic energies $\varepsilon_{\pm}=\frac{(\varepsilon_{i}^{\text{S}}+\varepsilon_{j}^{\text{M}})}{2}\pm\sqrt{\frac{\Delta^2+(\varepsilon_{i}^{\text{S}}-\varepsilon_{j}^{\text{M}})^2}{4}}$ are those shown in  \fig{fig:fig2}b. The probability of tunneling at a crossing time 
For a time dependent electric field $E(t)$, these distortions of the electronic structure will change in a dynamic fashion as the laser field develops. For strong ultrafast laser fields the system responds to the instantaneous value of the electric field as opposed to the pulse envelope. Thus, the Stark distortions due to $E(t)$ can be thought as ones in which the eigenstates of the system follow the instantaneous value of $E(t)$.  In this context, the effectiveness of a given crossing for interfacial charge transfer at a crossing time $t_{\text{crossing}}$  can be rationalized through  Landau-Zener tunneling probability $P_{\text{LZ}}=1-e^{-\beta}$, where 
 \begin{equation}
\label{eq:beta}
\beta=\frac{\pi\Delta^2}{2\hbar \left\vert \frac{d}{dt}\left(\varepsilon_i^{\text{S}}[E(t)]-\varepsilon_j^{\text{M}}[E(t)]\right)\right\vert_{t=t_{\text{crossing}}}}.
\end{equation}  
For strong laser fields, the Stark shifted energies vary linearly with the electric field such that $\frac{d\varepsilon_i^\alpha}{dt}\propto \frac{dE(t)}{dt}$. Thus the effectiveness of a crossing to open interfacial channels for electron transport 
increases with the energy gap at the crossing $\Delta$ and by reducing the laser frequency.

\subsection{Quantum Dynamics Simulations}

\subsubsection{Applicability of SCELI}
\begin{figure}[htb]
    \centering
\includegraphics[width=0.5\textwidth]{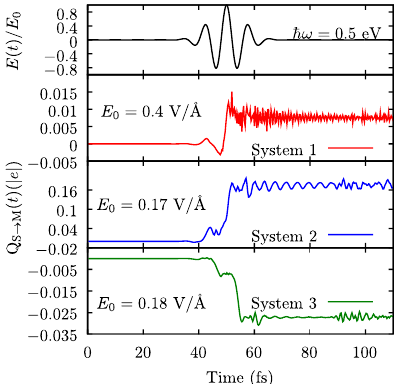}
\caption{Interfacial electron transfer induced by SCELI on the molecule-semiconductor interfaces. As the laser pulse is turned on, it opens quantum tunneling channels for electron transfer through Stark shifts. Here, $E_0$ is the amplitude of the laser with phase $\phi=0$ and central photon energy $\hbar\omega=0.5$ eV. {The field amplitudes employed in each case are chosen in regimes of the laser-matter interaction where the SCELI effect is appreciable for the particular interface.} The quantity $Q_{\text{S}\to\text{M}}(t)> 0$ when charge flows from S$\to$M, and negative when M$\to$S charge transfer is favored.\label{fig:fig3}}
\end{figure}

To demonstrate that SCELI is widely applicable in molecule-semiconductor heterojunctions, we propagate the quantum dynamics of three model heterojunctions with the energy lineups schematically shown in Fig. \ref{fig:fig1}b in the presence of few-cycle lasers $E(t)$ of the form in \eq{eq:laser}.  The simulations include all effects due to radiation-matter interactions including Stark effects and possible multiphoton transitions. However, the multiphoton transitions are suppressed by employing a laser frequency that is far detuned from electronic transitions in the problem. 

In system 1, the molecular orbitals are placed midgap. The energy difference between the conduction band (CB) edge and the LUMO ($\ket{\text{e}}$) is taken to be equal to that between the valence band (VB) edge and the HOMO ($\ket{\text{g}}$). In system 2, the VB and $\ket{\text{g}}$ spectrally overlap, while in system 3 the overlap is between $\ket{\text{e}}$ and the CB. These energy lineups are chosen to represent possible molecule-semiconductor interfaces. 

The Stark response of the molecule is governed by the transition dipole $\mu_\text{eg}$ and by permanent dipoles of the ground $\mu_\text{gg}$ and excited state $\mu_\text{ee}$. 
For definitiveness, consider first a molecule with an oscillator strength $f_{\text{OS}}=0.5$, a distance between molecular sites $a_{\text{M}}=3.38$ \AA, and no permanent dipoles ($\mu_\text{gg}= \mu_\text{ee}= 0$). \Fig{fig:fig3} show the electron transfer dynamics for the three systems generated by a few-cycle laser pulse of central frequency $\hbar\omega=0.5$ eV. The laser field onsets charge transfer $Q_{\text{S}\to\text{M}}(t)$ that terminates once the pulse is off. As discussed in \stn{stn:SCELI}, this is because only during the laser pulse the light-matter interaction creates transient resonances between the energy levels of the semiconductor's bands and the molecular orbitals that open quantum tunneling channels for interfacial charge transfer. 

Note that SCELI can be employed to generate charge transfer in all three heterojunctions, demonstrating the generality and versatility of the strategy. The energy level alignment in System 2 favors S$\to$M charge transfer due to transient resonances between VB levels and the molecular LUMO. Similarly, System 3 favors M$\to$S charge transfer due to transient resonances between CB levels and the HOMO. By contrast, the level alignment in System 1 does not favor M$\to$S or S$\to$M charge transfer. The net charge transfer that is observed from S$\to$M is because {of the larger available charge  in the semiconductor.}   The reduced magnitude of the SCELI effect in this case is due to the symmetry in the energy level alignment. 

Note that the control of the electron dynamics is due to the subcycle structure of light, leading to ultrafast electron transfer across the interface. In fact, the dynamics of the charge transfer bursts follows the peak structure of the electric field. 
Positive peaks in the electric field of light open S $\to$ M tunneling channels, while negative peaks  lead to M $\to$ S charge transfer. {Interestingly, in system 3 the $\text{S} \to \text{M}$  charge transfer is suppressed because the large energy difference between the   VB and the LUMO which prevents effective  $\text{S} \to \text{M}$ crossings for $E_0=0.18$ V/\AA. For this reason, the charge dynamics only shows the negative peaks of the electric field in this case. }

{After the pulse, the systems show small persistent oscillations in the transferred charge that arise due to electronic coherences between the molecular orbitals and the semiconductor states that are created during the light-matter interaction. For this reason,  these oscillations have frequencies that depend on the energy level alignment. System 1 shows the highest frequency oscillations as they involve superpositions between the VB and LUMO levels that are separated in energy by $\sim 4$ eV. The oscillations in system 2 (or 3) are slower as they involve superpositions between the VB and LUMO (or CB and HOMO) that are now separated by $\sim 1$ eV. While SCELI is robust to decoherence\cite{Antonio_2018,Antonio_2020}, these remnant oscillations not essential for SCELI are expected to rapidly decay due to electron-nuclear interactions\cite{doi:10.1021/acs.jpclett.7b03322}.}

\subsubsection{Dependence on laser parameters}

\begin{figure*}[htb]
\centering
\includegraphics[width=1.0\textwidth]{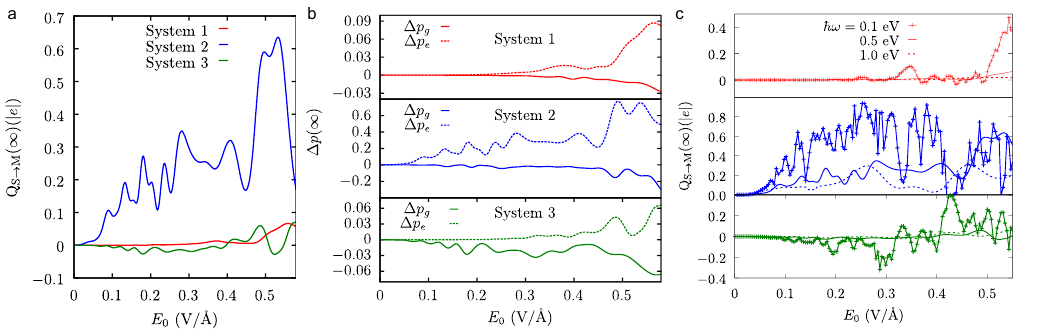}
\caption{Net interfacial electron transfer  on the molecule-semiconductor interfaces of \fig{fig:fig1}b induced by few cycle lasers. (a) Net charge transfer $Q_{\text{S}\to\text{M}}(\infty)$ and (b) associated changes in the molecular orbital population ($\Delta p_{\text{g}}$: HOMO; $\Delta p_{\text{e}}$: LUMO) as a function of laser amplitude $E_0$ ($\hbar \omega$=0.5 eV, $\phi=0$). (c) Net charge transfer for different laser frequency $\hbar \omega$ using pulses with the  same number of cycles ($\tau=\sqrt{2}\pi/\omega$) as a function of the laser amplitude $E_0$. \label{fig:fig4}}
\end{figure*}

\Fig{fig:fig4}a shows the dependence of SCELI on the laser amplitude $E_0$ for the three model heterojunctions. The figure shows the net charge transfer in the system after the laser pulse. In all cases there is a threshold laser amplitude $E_0$ for the onset of SCELI that arises due to the need for a sufficiently large laser amplitude for the Stark shifts to generate nontrivial crossings between the energy levels of molecule and semiconductor.  For this reason, the specific $E_0$ for which $Q_{\text{S}\to\text{M}}\neq 0$  depends on the energy-level alignment. The threshold amplitude is the largest for System 1, where the largest Stark shifts are needed to onset SCELI.  Once this threshold is reached, $Q_{\text{S}\to\text{M}}(\infty)$ generally increases with $E_0$ as the field is able to sample a larger number of transient resonances that lead to interfacial charge transfer. The SCELI effect is the largest for System 2 because this is where charge is mainly transferred from the semiconductor to molecule and the semiconductor has the largest available charge. 
 
\Fig{fig:fig4}b shows the net change of population in the molecular orbitals during SCELI. In System 1, SCELI is mostly due to population exchange from the VB of S to the LUMO. For large $E_0$, the HOMO$\to$S charge transfer also plays a role.  In system 2, the effect is mostly due to S$\to$ LUMO charge transfer. {In this case,  $Q_{\text{S}\to\text{M}}(\infty)$ shows an abrupt drop around $E_0=0.6$ V/\AA.
Contrasting Fig. \ref{fig:fig4}a and b we see that at such field amplitude the population of the HOMO and LUMO are both depleted with respect to SCELI dynamics at $\sim 0.5$ V/\AA. Thus, in this case, population of the HOMO and population previously deposited into the LUMO, SCELI their way into the semiconductor partially decreasing the overall charge transfer.} 
In system 3, for $E_0< 0.35$ V/\AA~ the effect is mostly due to HOMO$\to$ S charge transfer. For larger amplitudes S$\to$  LUMO charge transfer also plays a role and leads to oscillations in the overall directionality of SCELI.

\Fig{fig:fig4}c illustrates the dependence of SCELI on the laser frequency for the three systems. {We choose the pulse width $\tau=\sqrt{2\pi}/\omega$ such that all pulses have the same number of cycles and identical shape. This is important for a fair comparison, since SCELI is controlled by the subcycle structure of light.} As shown, decreasing the laser frequency $\hbar\omega$  increases the amount of charge that can be transferred through the SCELI mechanism.  This is because lowering the frequency makes the Landau-Zener transitions between diabatic levels more effective (see \eq{eq:beta}) since they increase the time range at which levels remain near resonance  thus enhancing the effectiveness of the tunneling mechanism.

\begin{figure}[htb]
\centering
\includegraphics[width=0.4\textwidth]{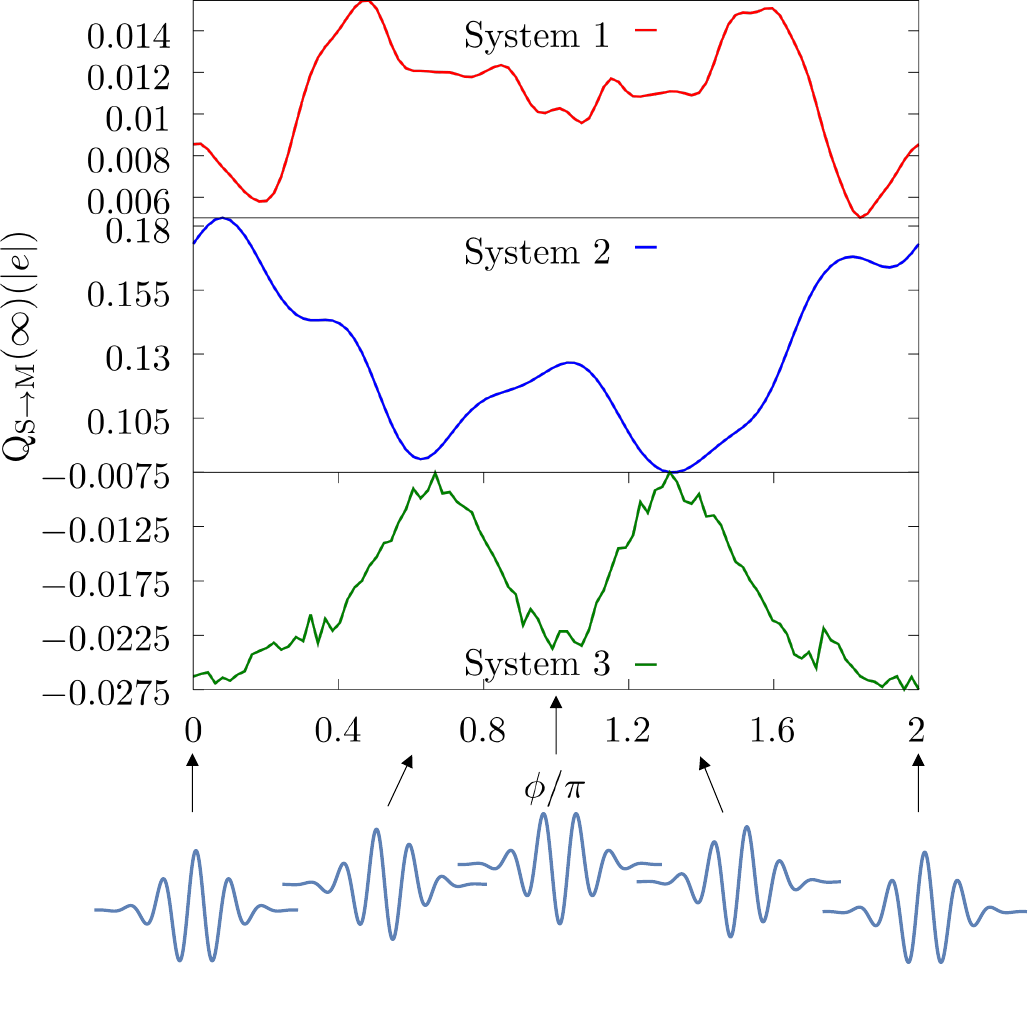}
\caption{Carrier envelope phase (CEP) dependence of SCELI. Net interfacial electron transfer  on the molecule-semiconductor interfaces of \fig{fig:fig1}b for lasers with varying $\phi$ (bottom panel). Modeling conditions are identical to those in \fig{fig:fig3} \label{fig:fig5}}
\end{figure}

{\Fig{fig:fig5} shows the influence of varying the CEP on SCELI for the three systems under the conditions in \fig{fig:fig3}. The bottom panel depicts how the shape of the laser field is modified by changing $\phi$. In general, varying the CEP changes the dynamics since SCELI reflects the subcycle structure of light. The mean effect as a function of CEP is non-zero since the molecule-semiconductor interface does not have an inversion symmetry.  Varying $\phi$ leads to a modulation of the effect by $\sim 50 \%$. }

\subsubsection{Dependence on molecular parameters}

\begin{figure}[htb]
\centering
\includegraphics[width=0.4\textwidth]{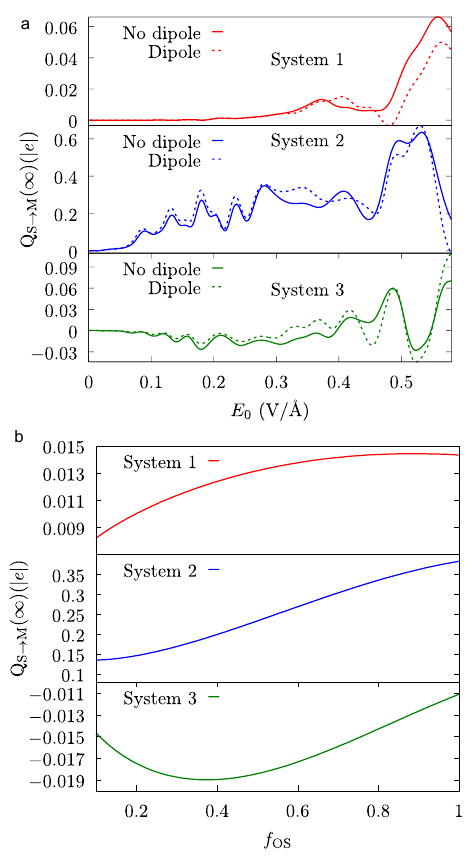}
\caption{Effect of the molecular dipolar character and oscillator strength on SCELI. The plots show the net charge transfer $Q_{\text{S}\to\text{M}}(\infty)$ across the heterojunction with the energy level alignments in Fig. \ref{fig:fig1}b using a laser pulse of $\phi=0$ and $\hbar\omega=0.5$ eV. (a) SCELI as a function of laser amplitude $E_0$ for a molecule with (dashed line, $\mu_{ee}=-\mu_{gg}$=4.8 D)  and without (solid lines) permanent dipoles with $f_{\text{OS}}=0.5$. (b) SCELI as a function of the molecular oscillator strength $f_{\text{OS}}$ as driven by a laser with amplitude of $E_0=$0.4, 0.17 and 0.18 V/\AA\ for the system 1 (red line), 2 (blue line), 3 (green line), respectively \label{fig:fig6}}
\end{figure}

We now investigate the effect of changing the molecular dipolar character and oscillator strength{, and the interfacial coupling,} on SCELI. \Fig{fig:fig6}a shows the net charge transfer $Q_{\text{S}\to\text{M}}(\infty)$ as a function of $E_0$ that is induced in the three systems when the molecule has a permanent dipole with $\mu_{ee}=-\mu_{gg}= 4.8$ D (dashed line) for which linear Stark effects emerge due to the permanent dipoles, and an oscillator strength $f_{\text{OS}}=0.5$. For clarity, we contrast these results with the case  in which $\mu_{ee}=\mu_{gg}=0$. As shown, the onset of SCELI, the magnitude of the effect, and its dependence on $E_0$ is only mildly affected by net molecular dipoles. This  is because the dominant effect that drives SCELI at these interfaces is the Stark response of the semiconductor.

\Fig{fig:fig6}b shows the effect of changing the oscillator strength $f_{\text{OS}}$ on SCELI for the three representative energy level alignments. For system 1 and 2, where S$\to$M charge transfer is favored, we observe an increase in $Q_{\text{S}\to\text{M}}(\infty)$ with $f_{\text{OS}}$ as this leads to crossings between VB levels of S with the LUMO for weaker field amplitudes. By contrast, in system 3, where SCELI is from $\text{M}\to\text{S}$, the dependence of SCELI on $f_{\text{OS}}$ is non-monotonic. The effect increases in magnitude up to a $f_{\text{OS}}=0.36$
where it has a maximum. The reduction of SCELI for $f_{\text{OS}}> 0.36$  is because quantum channels for $\text{S}\to\text{M}$ are opened as a consequence of the increase in the Stark effects with $f_{\text{OS}}$.

\begin{figure}[htb]
\centering
\includegraphics[width=0.4\textwidth]{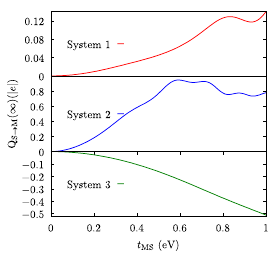}
\caption{Dependence of SCELI on the interfacial tight-binding coupling $t_{\text{MS}}$. Modeling conditions are identical to those
in Fig. 3\label{fig:fig7}}
\end{figure}

{
\Fig{fig:fig7} illustrates the dependence of the effect on the strength of the interfacial tight-binding coupling $t_\text{MS}$. As shown, generally increasing $t_{\text{MS}}$  enhances the SCELI magnitude as it enhances the tunneling  probability across the interface. Strong electronic coupling between the molecule and semiconductor is thus desirable for SCELI.  }

\subsection{SCELI vs. resonant routes for electron transfer}

A traditional way to induce electron transfer between a molecule and a semiconductor
is to resonantly photoexcite the semiconductor and allow the photoexcited carriers to transfer to a molecule that has empty energy levels that are resonant with the CB\cite{doi:10.1021/acs.chemrev.8b00392,WANG2017165}. One advantage of SCELI is that it can induce charge transfer in molecule-semiconductor junctions for which its energy lineup does not naturally allow interfacial charge transfer through resonant photoexcitation, such as in system 1.  
In circumstances where the energy-level aligment between semiconductor and molecule does allow for charge exchange upon resonant photoexcitations of S (such as system 2 and 3), an advantage of SCELI is that the electron transfer process is faster as it happens during the laser pulse and, moreover, it is controlled by the subcycle structure of light (\emph{cf.} \figs{fig:fig3}{fig:fig5}). By contrast in resonant photoexcitation the laser itself does not lead to charge transfer. The charge transfer occurs after photoexcitation and its rate is controlled by the electronic coupling among molecular and semiconductor levels. Thus SCELI mechanisms for interfacial charge transfer are faster than conventional resonant routes and can be employed to circumvent unfavorable energy level alignments.

\section{Conclusions}
\label{stn:conclusions}

In conclusion, we have demonstrated interfacial charge transfer at molecule-semiconductor interfaces induced by the SCELI mechanism. SCELI  is based on creating transient resonances between molecular and semiconductor levels through Stark shifts generated by non-resonant few cycle lasers of intermediate intensity. These transient resonances open tunneling channels for electron transfer that can be used to laser control the electron dynamics at the interface. The effect was exemplified in three representative energy alignments for the interface shown in Fig. \ref{fig:fig1}b. SCELI is robust to the energy level alignment but is most effective when there is a large density of states in donor levels. That is, when the charge exchange is from the valence band of the semiconductor to the molecular excited state.

Further, the effect is seen to be relatively insensitive to the molecular oscillator strength and dipolar character as it is the Stark response of the semiconductor what dominates the SCELI dynamics. Further, SCELI can be employed to induce interfacial charge transfer when resonant routes are not available and in timescales faster than those offered by resonant photoexcitation. 

{The SCELI effect can be generated with lasers of intermediate intensity $\lesssim  5 \times 10^{12}$ W/cm$^2$ (amplitudes $E_0 \lesssim 0.6$ V/\AA) that  distort the electronic structure of matter through Stark shifts but that do not generate  ionization or molecular dissociation. The latter effects emerge at even stronger laser intensities $>10^{13}$ W/cm$^2$ \cite{doi:10.1021/jp984543v} and can be suppressed by further detuning the laser frequency from transitions in the molecule and the semiconductor. Thus there is a well-defined regime where SCELI can be isolated from competing phenomena. }

Contrasting the results of SCELI at molecule-semiconductor interfaces with those for semiconductor-semiconductor interfaces \citep{Antonio_2018,Antonio_2020,Antonio_2020_2} shows that the underlying mechanism remains intact underscoring the generality of the scheme to control electron dynamics at interfaces.

\begin{acknowledgments}
This material is based upon work supported by the National Science Foundation under Grant No. CHE-2102386.
\end{acknowledgments}

\bibliography{references}

\end{document}